\documentclass[twocolumn, 10pt, amssymb, aps, physrev, longbibliography]{revtex4-2}

\usepackage{graphicx}
\graphicspath{ {graphics/} }
\usepackage{amsmath}
\usepackage{siunitx}

\usepackage[utf8]{inputenc}

\usepackage[colorlinks=true, linkcolor=black, citecolor=black, urlcolor=black, unicode=true]{hyperref}

\newcommand\I{\mathrm{i}}
\newcommand\e{\mathrm{e}}
\newcommand\di{\mathrm{d}}
\newcommand\hc{\mathrm{h.c.}}
\DeclareMathOperator{\Tr}{Tr}
\let\vec\mathbf

\begin{document}

\title{Steering edge currents through a Floquet topological insulator}

\author{Helena Drüeke}
\email{helena.drueeke@uni-rostock.de}
\author{Marcus Meschede}
\email{marcus.meschede@uni-rostock.de}
\author{Dieter Bauer}
\email{dieter.bauer@uni-rostock.de}
\affiliation{Institute of Physics, University of Rostock, 18051 Rostock, Germany}

\begin{abstract}
Periodic driving may cause topologically protected, chiral transport along edges of a 2D lattice that, without driving, would be topologically trivial.
We study what happens if one adds a different on-site potential along the diagonal of such a 2D grid.
In addition to the usual bulk and edge states, the system then also exhibits doublon states, analogous to two interacting particles in one dimension.
A particle initially located at an edge propagates along the system's boundary.
Its wave function splits when it hits the diagonal and continues propagating simultaneously along the edge and the diagonal.
The strength of the diagonal potential determines the ratio between both parts.
We show that for specific values of the diagonal potential, hopping onto the diagonal is prohibited so that the system effectively separates into two triangular lattices.
For other values of the diagonal potential, we find a temporal delay between the two contributions traveling around and through the system.
This behavior could enable the steering of topologically protected transport of light along the edges and through the bulk of laser-inscribed photonic waveguide arrays.
\end{abstract}

\maketitle

\section{Introduction}
\label{sec:introduction}
Topological insulators~\cite{hasan_colloquium_2010, qi_topological_2011} exhibit an insulating bulk and conducting edges.
Floquet topological insulators~\cite{oka_photovoltaic_2009, kitagawa_topological_2010, lindner_floquet_2011} are synthetic systems.
Periodic driving causes their behavior as topological insulators.
Therefore, time becomes an additional system dimension, and modifications of the driving scheme are a practical parameter for tuning Floquet topological insulators.
They are realized experimentally with cold atoms in optical lattices~\cite{jotzu_experimental_2014, wintersperger_realization_2020, cooper_topological_2019}, with photonic platforms~\cite{segev_topological_2021, ozawa_topological_2019-1, kirsch_nonlinear_2021, ivanov_topological_2021, rechtsman_photonic_2013, hafezi_imaging_2013}, or even for acoustic waves~\cite{peng_experimental_2016, fleury_floquet_2016}.
Theoretically, they are studied in 1D~\cite{asboth_chiral_2014, dal_lago_floquet_2015}, 2D~\cite{graf_bulkedge_2018, unal_hopf_2019}, and systematized in a periodic table~\cite{roy_periodic_2017}.
For a recent theoretical review, see~\cite{rudner_band_2020}.

Rudner {\em et al.}~\cite{rudner_anomalous_2013} proposed a Floquet topological insulator on a bipartite square lattice.
They also derived a topological invariant for that system.
In our publication, we investigate a modified version of Rudner's system. Instead of global, alternating on-site potentials on the two sublattices, we introduce an on-site potential along the diagonal of the lattice.
We examine how the diagonal potential can cut the system in half, fully or partially.
Without it, a topologically protected edge current can flow along the system's outer edge.
The addition of a diagonal potential introduces an "inner" edge.
Depending on the exact choice of parameters, a portion or the whole edge state may flow along this inner edge instead of the outer edge.

The diagonal potential leads to some eigenstates located (almost) exclusively along the diagonal.
We relate these eigenstates to doublon physics.
There, two interacting particles on, for example, a one-dimensional~\cite{valiente_lattice_2010, di_liberto_two-body_2016, marques_topological_2018, azcona_doublons_2021} or two-dimensional~\cite{salerno_interaction-induced_2020} lattice may form a bound pair, a so-called doublon state.
Doublons are stable due to their energetic separation from other states or because of topology, characterized by a differing topological invariant.
They can arise from repulsively as well as from attractively interacting particle pairs.

In order to prove that our system is topologically non-trivial, we need a topological invariant.
The usual Chern number is insufficient to determine the existence of chiral edge states in Floquet systems.
The Chern number of an energy band gives the difference between the numbers of chiral edge modes entering it from above and below.
In a static system, the energy spectrum is bound from below, i.e., no edge modes can exist below the lowest-energy band.
The number of edge states in any band gap equals the sum of the Chern numbers of all bands below it.
However, in Floquet systems, due to their temporal periodicity, the energy becomes a quasi-energy, which is also periodic.
Edge modes can loop around from below the lowest to above the highest band.
This allows chiral edge states to exist in a Floquet system in which all bands have Chern number 0.
Rudner {\em et al.}~\cite{rudner_anomalous_2013} derived a winding number as a topological invariant for Floquet systems.
In contrast to the Chern number, which only depends on projectors onto Floquet bands, the winding number considers the whole time evolution throughout the driving cycle.

This publication is structured as follows:
Sec.~\ref{sec:system} contains a description of the system under study, Sec.~\ref{sec:floquet} its Floquet eigenstates, Sec.~\ref{sec:bandstructure} its band structure, and Sec.~\ref{sec:results_time} the temporal evolution of states.
We give a conclusion in Sec.~\ref{sec:conclusion}.
The appendix contains derivations of the hopping probability (App.~\ref{sec:derivation_probability}) and the topological invariant (App.~\ref{sec:topological_invariant}), as well as the algorithm for tracking states through (avoided) crossings (App.~\ref{sec:tracking}).

\section{System}
\label{sec:system}
\begin{figure}[htbp]
\centering
\includegraphics{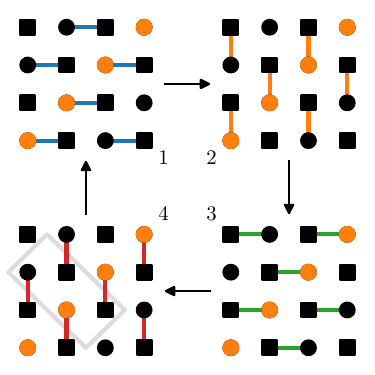}
\caption{
Sketch of the driving scheme for a $4 \times 4$ system.
The dots (squares) indicate the lattice sites on sublattice A (B).
The orange dots are the lattice sites on the diagonal (with a modified on-site potential).
The four panels show the different hoppings in the four phases of the driving cycle as (blue, orange, green, and red) connecting lines.
The gray rectangle indicates the unit cell used for calculating the band structure in Sec.~\ref{sec:bandstructure}.
}
\label{fig:driving_scheme}
\end{figure}
We investigate an $N \times N$ square-lattice system.
The lattice sites $s_{i, j}$ are numbered by horizontal and vertical indices $i, j \in \{1, 2, \ldots, N\}$.
Index $i$ increases from left to right, and $j$ increases from bottom to top.
Sublattice A contains all sites with even sum $i + j$, and sublattice B those with odd $i + j$.
The periodic driving scheme of this Floquet system is sketched in \autoref{fig:driving_scheme}.
We modify the potentials of the sites $s_{i,\, i}$ along the diagonal (as shown in \autoref{fig:driving_scheme}), which we set to $V_\mathrm{dia}$.
All other on-site potentials are $0$.

\subsection{Hamiltonians}
We will now introduce a mathematical description of the system and its parameters.
The hopping amplitudes and potentials are described by the step-wise constant Hamiltonian
\begin{equation}
\hat{H}(t) =
\begin{cases}
\hat{H}_1 & 0 < t \le T/4 \\
\hat{H}_2 & T/4 < t \le T/2 \\
\hat{H}_3 & T/2 < t \le 3T/4 \\
\hat{H}_4 & 3T/4 < t \le T
\end{cases}
\label{equ:Hamiltonian}
\end{equation}
with
\begin{equation}
\hat{H}_k =
\hat{H}_\mathrm{dia} +
J \sum_{\substack{i,j = 1 \\ i + j \ \mathrm{even}}}^N
\hat{h}_k(i, j) + \mathrm{h.c.},
\end{equation}
\begin{equation}
\hat{H}_\mathrm{dia} = V_\mathrm{dia} \sum_{i = 1}^N |i, i\rangle \langle i, i|,
\end{equation}
\begin{equation}
\begin{aligned}
\hat{h}_1(i, j) &=
|i, j\rangle \langle i + 1, j|,
\ &\
\hat{h}_2(i, j) &=
|i, j\rangle \langle i, j + 1|,
\\
\hat{h}_3(i, j) &=
|i, j\rangle \langle i - 1, j|,
\ &\
\hat{h}_4(i, j) &=
|i, j\rangle \langle i, j - 1|,
\end{aligned}
\end{equation}
period $T$, and hopping amplitude $J$.

\subsection{Hopping probabilities}
\label{sec:hopping_probabilities}
In each of the four phases of the driving scheme, each lattice site connects to (at most) one neighboring lattice site.
Therefore, the Hamiltonian of the whole system breaks down into several two-state Hamiltonians, each describing the interaction between connected sites.
In appendix \ref{sec:derivation_probability}, we derive the hopping probability between two connected sites
\begin{equation}
p(t, V) = \frac{4 J^2}{V^2 + 4 J^2} \sin^2\left(\frac{\sqrt{V^2 + 4 J^2}}{2} t\right)
\end{equation}
as a function of the potential difference $V$ between the sites and the duration $t$ of the respective driving phase.
Without a potential (i.e., between the off-diagonal sites), we want the particles to hop completely during each of the four phases, $p\left(\frac{T}{4}, 0\right) = \sin^2\left(\frac{J T}{4}\right) := 1$.
We set $J = 1$
and choose $T = \frac{2 \pi}{J} = 2 \pi$, such that in the absence of on-site potentials, the hopping probability between two connected sites during each phase is exactly one.
With on-site potential $V_\mathrm{dia}$ along the diagonal, the hopping probability between connected diagonal and off-diagonal sites (and vice versa) becomes
\begin{equation}
p(V_\mathrm{dia})
= p\left(\frac{T}{4}, V_\mathrm{dia}\right)
= \frac{4}{V_\mathrm{dia}^2 + 4} \sin^2\left(\frac{\pi}{4} \sqrt{V_\mathrm{dia}^2 + 4}\right),
\label{eq:prob_pot}
\end{equation}
shown in Fig.~\ref{fig:localization}a.
The prefactor $4/(V_\mathrm{dia}^2 + 4)$
leads to a decrease in the hopping probability at higher diagonal potentials.
As expected, the particle cannot overcome the energetic separation between off-diagonal and diagonal sites at high diagonal potentials.
But the probability for hopping onto the diagonal is also vanishing  at finite diagonal potentials
\begin{equation}
V^{(n)}_\mathrm{dia, 0} = \pm 2 \sqrt{4 n^2 - 1}
\label{eq:prob_zeros}
\end{equation}
for $n \in \mathbb{N}$, determined by the zeros of the $\sin^2$-term of \eqref{eq:prob_pot}.
At these specific potentials, the duration of each phase corresponds to one complete Rabi period.
Therefore, at $V^{(n)}_\mathrm{dia, 0}$, a particle moves from its starting site to the connected neighboring site and back $n$ times, returning to the starting site at the end of the phase.
The first zero is at a remarkably low diagonal potential $V^{(1)}_\mathrm{dia, 0} = 2 \sqrt{3}$.

\subsection{Time Evolution Operator}
\label{subsec:time_evolution_operator}
The time evolution operator (in units where $\hbar=1$)
\begin{equation}
\hat{U}(t)
=\mathcal{T} \exp \left(-\I \int_0^t \di t' \ \hat{H}(t') \right),
\end{equation}
where $\mathcal{T}$ is the time-ordering operator, describes the system's temporal evolution.
For a complete driving cycle consisting of the four discrete phases ($t = T$), the time evolution operator reads
\begin{equation}
\hat{U}(T)
= \e^{c \hat{H}_4} \e^{c \hat{H}_3} \e^{c \hat{H}_2} \e^{c \hat{H}_1}  ,
\end{equation}
with $c = \frac{T}{4 \I}$.
Solving the equation
\begin{equation}
\hat{U}(T) \psi_\mathrm{F} = \lambda_\mathrm{F} \psi_\mathrm{F}
\end{equation}
gives the Floquet eigenstates $\psi_\mathrm{F}$, and
the Floquet energies $\varepsilon_\mathrm{F}$ are calculated from the eigenvalues $\lambda_\mathrm{F} = \exp(- \I \varepsilon_\mathrm{F} T)$.
By following the states (not just their energies) as a function of $V_\mathrm{dia}$, we can distinguish state crossings from avoided crossings.
Appendix \ref{sec:tracking} explains the details of the state-tracking algorithm.

\section{Floquet states}
\label{sec:floquet}
\begin{figure}[htbp]
\centering
\includegraphics{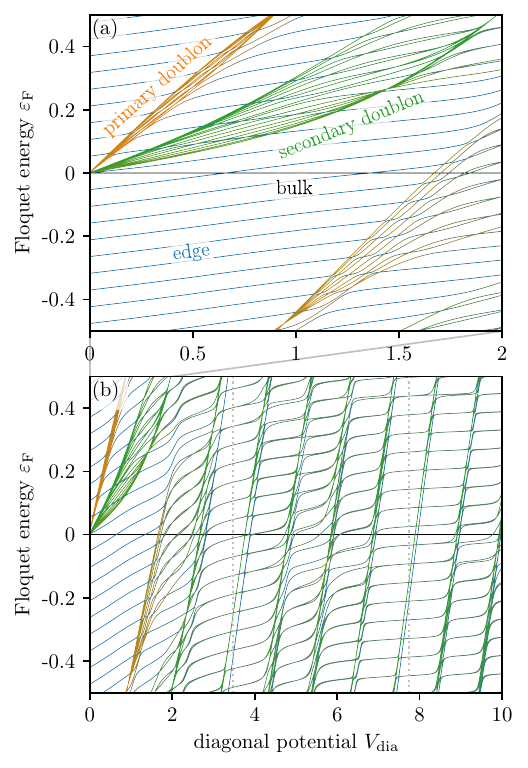}
\caption{
Floquet energies $\varepsilon_\mathrm{F}$ for the $10 \times 10$ system as a function of the diagonal potential $V_\mathrm{dia}$.
a) The line colors indicate the types of states: bulk states (black), edge states (blue), primary doublon states (orange), and secondary doublon states (green).
b) The dotted gray lines indicate the zeros $V^{(1)}_\mathrm{dia, 0}$, $V^{(2)}_\mathrm{dia, 0}$ of the hopping probability $p(V_\mathrm{dia})$.
}
\label{fig:energies}
\end{figure}

Without a diagonal potential, the $N \times N$ system contains $(N - 1)^2$ bulk states at Floquet energy $\varepsilon_\mathrm{F} = 0$ and $2 N - 1$ edge states at equally spaced Floquet energies $\varepsilon_\mathrm{F} = m / (2 N - 1)$, $m \in \mathbb{Z}$, $|m| < N$.
At non-vanishing diagonal potential, some of the bulk states superpose to form doublon states, resulting in $(N - 3) (N - 2)$ remaining bulk states, $N - 1$ primary doublon states, and $2 (N - 2)$ secondary doublon states.
Particles in a primary doublon state hop on and off the modified diagonal twice during a complete driving cycle.
They hop on and off the diagonal once in a secondary doublon state.

We call these diagonal states \emph{doublons}, in analogy to doublons in one-dimensional, interacting systems~\cite{valiente_lattice_2010, di_liberto_two-body_2016, marques_topological_2018, salerno_interaction-induced_2020, azcona_doublons_2021}.
A system of two interacting particles on a one-dimensional lattice (e.g., an SSH chain~\cite{su_solitons_1979}) can be mapped to a single particle on a two-dimensional lattice. The $x$- and $y$-coordinates of the 2D particle correspond to the position of the first and the second particle on the 1D chain, respectively.
Movements of the first (second) particle then amount to horizontal (vertical) hoppings.
Their interaction $V(|x-y|)$ is captured by an on-site potential in 2D.
A local interaction in 1D, through which the particles only affect each other if they are on the same site, corresponds to the diagonal potential along $x=y$ used in this study.
Longer-range interactions would require non-vanishing on-site potentials on not only the main but also the secondary, tertiary, etc.\ diagonal, depending on the maximum interaction distance.
However, in our Floquet system, the mapping between the two-particle-1D and the one-particle-2D case is more involved and not that intuitive because the simple driving scheme in 2D corresponds to intricate, time-dependent, non-local interactions of two particles in 1D.

Figure \ref{fig:energies} shows the Floquet energies $\varepsilon_\mathrm{F}$ as a function of the diagonal potential $V_\mathrm{dia}$.
The four different types of states are indicated in \autoref{fig:energies}a:
The diagonal potential does not influence the bulk states' energies because the bulk states remain located on sites whose potentials are zero in each step of the driving cycle.
The edge states cross the diagonal in the system's bottom left and top right corners.
Therefore, their energies increase with increasing diagonal potential.
The doublons cross the diagonal as well, which leads to increasing Floquet energies.
The primary doublons' energies increase faster because they cross the diagonal twice during a driving cycle, unlike the secondary doublons, which only cross it once.

At higher diagonal potentials $V_\mathrm{dia} \gtrsim 2$ (shown in \autoref{fig:energies}b), three types of states exist:
The bulk states' energies remain zero, while the edge states' energies increase slightly with diagonal potential.
For the Floquet states located on the diagonal of the system, we find $\varepsilon_F \approx V_\mathrm{dia}$.
All states form pairs with similar energies.
These pairs degenerate at the zeros of the hopping probability, and we find superpositions of the states confined to either the top left or bottom right triangular half of the system.

\section{Band structure}
\label{sec:bandstructure}
\begin{figure}
\centering
\includegraphics{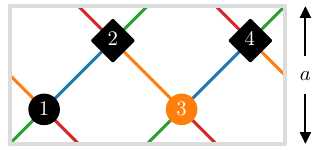}
\caption{
Unit cell of the $\ang{45}$-rotated system as outlined in light gray in panel 4 of Fig.~\ref{fig:driving_scheme}.
The system is infinitely extended in the vertical direction.
The height of the unit cell, $a$, is marked.
Periodic boundaries are employed horizontally.
The orange circle (3) marks the diagonal.
The hoppings of all four phases are shown in the same colors as in Fig.~\ref{fig:driving_scheme}.
}
\label{fig:sketch_dia}
\end{figure}

In order to calculate a band structure, we must have a system that is infinitely extended in at least one direction.
We turn our system by $\ang{45}$ and obtain an infinite strip with the former diagonal in the middle.
The narrowest viable strip is four sites wide, as shown in Fig.~\ref{fig:sketch_dia}.
Wider strips would only host more bulk states with identical energies.
The unit cell is repeated infinitely in the vertical direction and numbered by an index $m$.
The sites are numbered $\alpha=1,2,3,4$.

We can write the Hamiltonians for the four phases in real space as
\begin{equation}
\hat{H}_i = V | m, 3 \rangle \langle m, 3 | + \left(J \sum_m \hat{h}_i(m) + \hc\right)
\end{equation}
with
\begin{equation}
\begin{aligned}
\hat{h}_1(m) &= | m, 1 \rangle \langle m, 2 | + | m, 3 \rangle \langle m, 4 | \\
\hat{h}_2(m) &= | m, 3 \rangle \langle m, 2 | + | m, 1 \rangle \langle m, 4 | \\
\hat{h}_3(m) &= | m + 1, 3 \rangle \langle m, 2 | + | m + 1, 1 \rangle \langle m, 4 | \\
\hat{h}_4(m) &= | m + 1, 1 \rangle \langle m, 2 | + | m + 1, 3 \rangle \langle m, 4 |.
\end{aligned}
\end{equation}

We transform the Hamiltonians to $k$-space by making the Bloch ansatz
\begin{equation}
| m, \alpha \rangle = \frac{a}{2\pi} \int_\mathrm{BZ} \di k \, \exp(-\I  k m a) \, | k, \alpha \rangle,
\end{equation}
where $a$ is the height of the unit cell (i.e., the distance from site $1$ in cell $m$ to site $1$ in cell $m + 1$).
We can identify
\begin{equation}
\hat{H}_i = \frac{a}{2\pi}\int_{\mathrm{BZ}} \di k \, | k \rangle \langle k | \left(V |3 \rangle \langle 3 | + J \left(\hat{h}_i(k) + \hc\right)\right)
\end{equation}
with
\begin{equation}
\begin{aligned}
\hat{h}_1(k) &= |1 \rangle \langle 2 | + | 3 \rangle \langle 4 | \\
\hat{h}_2(k) &= |3 \rangle \langle 2 | + | 1 \rangle \langle 4 | \\
\hat{h}_3(k) &= \exp(\I k a) \left(| 3 \rangle \langle 2 | + | 1 \rangle \langle 4 |\right) \\
\hat{h}_4(k) &= \exp(\I k a) \left(| 1 \rangle \langle 2 | + | 3 \rangle \langle 4 |\right).
\end{aligned}
\end{equation}

\begin{figure}
\centering
\includegraphics{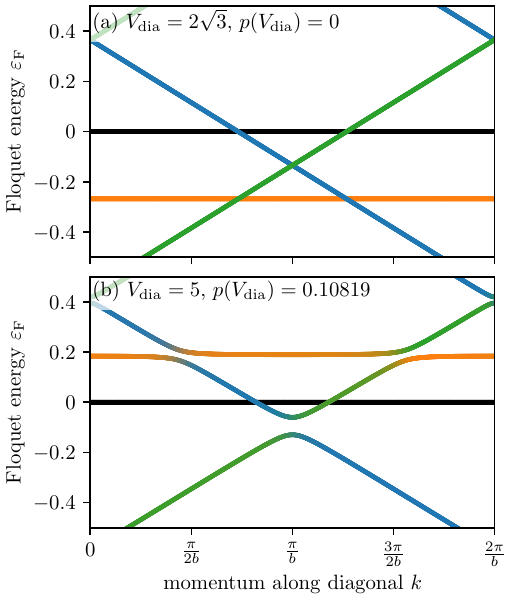}
\caption{
Bandstructure of a $4$-site wide diagonal strip at different diagonal potentials $V_\mathrm{dia}$.
The bulk states (shown in black) form a flat band at $\varepsilon_\mathrm{F} = 0$, irrespective of the diagonal potential.
a) $V_\mathrm{dia}=2\sqrt{3}$, i.e., $p(V_\mathrm{dia})=0$, the primary diagonal states form a flat band (orange and the secondary diagonal states form diagonal bands (green and blue).
Transferring between bands is impossible.
b) $V_\mathrm{dia}=5$, i.e., $p(V_\mathrm{dia})=0.10819$, the crossings between (primary and secondary) diagonal states become avoided crossings.
}
\label{fig:bands}
\end{figure}

We used these Hamiltonians to calculate the band structures shown in Fig.~\ref{fig:bands}.
At diagonal potentials $V_\mathrm{dia} = V^{(n)}_\mathrm{dia, 0}$ (Fig.~\ref{fig:bands}a), the bandstructure consists of four separate bands:
The flat band at $\varepsilon_\mathrm{F} = 0$ (shown in black) consists of stationary bulk states, and the other flat band (orange) consists of primary diagonal states, which are also stationary.
The two diagonal bands (green and blue) consist of secondary diagonal states.
Their slope indicates the movement of these states along the diagonal.
The bulk band remains unchanged at other diagonal potentials (Fig.~\ref{fig:bands}b).
The three other bands, however, connect as their crossings turn into avoided crossings.

We employed periodic boundary conditions on the left and right edges of the unit cell when calculating the band structures.
The full system is not periodic in this direction.
It only contains a single modified diagonal, not a periodically repeated one.
For the calculation of the band structure, however, this periodicity (the connection between sites 1 and 4 in Fig.~\ref{fig:sketch_dia}) is necessary.
Without it, the left and right edges of the unit cell would become edges of the system, hosting edge states (two additional diagonal bands).
Such edges, which would lie in diagonal direction in the non-rotated system, do not exist in the full (square) system.
It only contains horizontal and vertical edges, whose edge states move with a different velocity than the diagonal ones.
These edge states would also result in diagonal bands but with a different slope.
The band structures in Fig.~\ref{fig:bands} do not contain these bands.
They are band structures of the bulk, showing stationary states and those moving along the diagonal.

A topological invariant for the system is derived in App.~\ref{sec:topological_invariant}.

\section{Temporal evolution}
\label{sec:results_time}
\begin{figure}[htbp]
\centering
\includegraphics{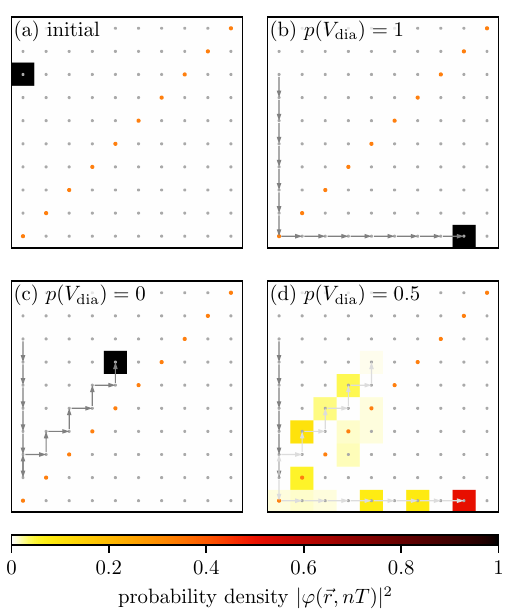}
\caption{
Temporal evolution of a localized edge state in the $10 \times 10$ system.
The circles mark the locations of the sites.
The orange ones are those on the diagonal.
a) Starting configuration ($n = 0$).
b) - d) State after evolution for $n = 8$ complete cycles.
The arrows sketch the path of the state during its evolution.
b) $V_\mathrm{dia}=0$, i.e., $p(V_\mathrm{dia})=1$, the state moves around the whole system and will return at its initial site after $2 N - 1 = 19$ cycles.
c) $V_\mathrm{dia}=2\sqrt{3}$, i.e., $p(V_\mathrm{dia})=0$, the state moves around the upper left triangle because the diagonal is insurmountable and will return after $2 N - 3 = 17$ cycles.
d) $V_\mathrm{dia}=1.59737$, i.e. $p(V_\mathrm{dia})=0.5$, the state splits.
The supplementary material~\cite{note:supplement} contains animations of these three cases for a $20 \times 20$ system.
}
\label{fig:temporal}
\end{figure}

By definition, the Floquet states $\psi_\mathrm{F}$ remain unchanged after a complete cycle's evolution, except for a phase factor. However, for an intuitive understanding of the driven quantum dynamics, it is instructive (and closer to experimental realizations on photonic platforms) to study the temporal evolution of (initially) localized states $\varphi(\vec{r}, t)$. We initialize these as one at a single site $\vec{r}_\mathrm{initial}$ and zero everywhere else.
Specifically, in \autoref{fig:temporal}, we look at a state starting on the left edge of the $10 \times 10$ system.

Without a diagonal potential and therefore perfect hopping $p(0) = 1$ between diagonal and off-diagonal sites (\autoref{fig:temporal}b), the state remains localized and moves counter-clockwise along the edges of the system, and returns to its origin after $2 N - 1$ complete driving cycles.
The state cannot cross the diagonal at the zeros of the hopping probability $p(V_\mathrm{dia}) = 0$.
Because it started above the diagonal, it will remain in the upper left triangular half of the system, moving along the edge of that triangle and returning to its origin after $2 N - 3$ cycles (\autoref{fig:temporal}c).
For all other potentials, which result in $0 < p(V_\mathrm{dia}) < 1$, the state splits at the bottom left corner of the system and delocalizes (\autoref{fig:temporal}d).

Most ($(N-3)(N-2)$) of the $(N-1)^2$ bulk states do not interact with any sites on the diagonal, and therefore their evolution is unaffected by $V_\mathrm{dia}$.
The other $3N-5$ bulk states, however, interact with the diagonal at least once and are split up for $0 < p(V_\mathrm{dia}) < 1$.
For $p(V_\mathrm{dia}) = 0$, $2N-3$ bulk states turn into edge states on the upper left and lower right triangles.
The $N-2$ bulk and $2$ edge states that start on the diagonal remain stationary for $p(V_\mathrm{dia}) = 0$.

\begin{figure}[htbp]
\centering
\includegraphics[width=\columnwidth]{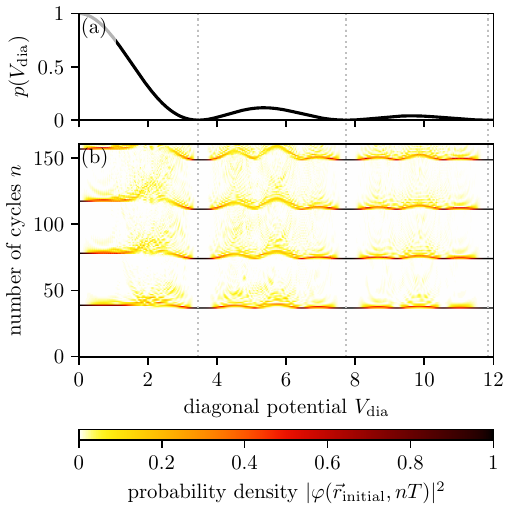}
\caption{
a) Hopping probability $p(V_\mathrm{dia})$ as a function of diagonal potential.
b) Probability density $|\varphi(\vec{r}_\mathrm{initial}, n T)|^2$ ($n \in \mathbb{N}$) at the starting location $\vec{r}_\mathrm{initial}$ for an edge state in a $20 \times 20$ system, as a function of time and diagonal potential.
The dotted gray lines indicate the zeros of $p(V_\mathrm{dia})$ in a), which coincide with the perfect earlier returns in b).
}
\label{fig:localization}
\end{figure}

While snapshots of the temporal evolution allow us to understand the general behavior, it is inconvenient to investigate them systematically for varying potentials.
Therefore, we record the probability density $|\varphi(\vec{r}_\mathrm{initial}, n T)|^2$ ($n \in \mathbb{N}$) at the starting location $\vec{r}_\mathrm{initial}$ after each driving cycle.
Because we initialize all states perfectly localized, $|\varphi(\vec{r}_\mathrm{initial}, 0)|^2 = 1$ for all states.
The bulk states return to their starting location after each cycle, and therefore $|\varphi_\mathrm{bulk}(\vec{r}_\mathrm{initial}, n T)|^2 = 1\ \forall n \in \mathbb{N}$.

As shown in \autoref{fig:localization}, without a diagonal potential, an edge state returns at its origin after $2 N - 1$ cycles.
At the zeros of $p(V_\mathrm{dia})$  \eqref{eq:prob_zeros},
the system is equivalent to one without sites on the diagonal.
The edge states remain confined to the upper left or bottom right triangle, which they started in, and return to their origins after $2 N - 3$ cycles.
Other on-site potentials $V_\mathrm{dia}$ lead to a splitting of the state.
Part of the wavefunction travels through the diagonal and along the square's edge, arriving after $2 N - 1$ cycles, while another part travels around the triangle and arrives after $2 N - 3$ cycles.
Due to the imperfect transfer, there are not two discrete arrival peaks.
Instead, the peaks disperse and interfere.

\section{Conclusion and outlook}
\label{sec:conclusion}
We characterized a Floquet topological insulator on a square lattice with varying diagonal on-site potential.
The addition of a diagonal potential causes the appearance of additional bands.
We call these states doublons, in analogy to doublons in systems of one-dimensional, interacting particles.
There are stationary doublons and those which perform directional motion along the diagonal.

We showed how fine-tuning the diagonal potential cuts the Floquet topological insulator into halves.
While a small diagonal potential is sufficient to disconnect the system completely, surprisingly, increasing the diagonal potential further will partially reconnect the system.
This partial cutting allows switching between the propagation along the edge and the diagonal.
In a photonic setting with laser-inscribed waveguides, intense laser light and nonlinearities can modify the diagonal potential (i.e., refractive index)~\cite{maczewsky_nonlinearity-induced_2020, kirsch_nonlinear_2021, ivanov_topological_2021}.
In that way, one may switch topologically protected light currents by light, rendering a photonic platform programmable instead of "hard-wired."

\section*{Acknowledgment}
This research and publication were supported by the Studienstiftung des deutschen Volkes (German Academic Scholarship Foundation) and the Deutsche Forschungsgemeinschaft (DFG, German Research Foundation), SFB 1477 "Light-Matter Interactions at Interfaces," project number 441234705.

\appendix
\section{Derivation of hopping probability}
\label{sec:derivation_probability}
The $2 \times 2$ Hamiltonian
\begin{equation}
\hat{H}
=
\begin{pmatrix}
V & J \\
J & 0
\end{pmatrix}
\end{equation}
describes the hopping between two sites with potential difference $V$ and hopping element $J$.
Its eigenstates
\begin{equation}
\varphi_{1,2}
=
\begin{pmatrix}
V \pm \sqrt{V^2 + 4 J^2} \\
2 J
\end{pmatrix}
\end{equation}
and corresponding eigenenergies
\begin{equation}
\varepsilon_{1,2} = \frac{1}{2} \left(V \pm \sqrt{V^2 + 4 J^2}\right)
\end{equation}
satisfy the time-independent Schrödinger equation
\begin{equation}
\hat{H} \varphi_{1,2} = \varepsilon_{1,2} \varphi_{1,2}.
\end{equation}
The solutions of the time-dependent Schrödinger equation
\begin{equation}
\hat{H} \psi(t) = \I \partial_t \psi(t)
\end{equation}
are superpositions of the eigenstates
\begin{equation}
\psi(t) =  a_1 \exp\left(-\I \varepsilon_1 t\right) \varphi_1 + a_2 \exp\left(-\I \varepsilon_2 t\right) \varphi_2.
\end{equation}
For a time-dependent state starting localized on the first lattice site,
\begin{equation}
\begin{pmatrix}
1 \\
0
\end{pmatrix}
=
\psi(0)
=
a_1 \varphi_1 + a_2 \varphi_2,
\end{equation}
we find
\begin{equation}
a_1 = -a_2 = \frac{1}{2 \sqrt{V^2 + 4 J^2}}.
\end{equation}
The hopping probability is the absolute square of the time-dependent state at the second lattice site,
\begin{align}
p(t, V) &= |\psi_2(t)|^2 \\
  &= \frac{4 J^2}{V^2 + 4 J^2} \sin^2\left(\frac{\sqrt{V^2 + 4 J^2}}{2} t\right).
\end{align}

\section{Topological Invariant}
\label{sec:topological_invariant}
In the following, we discuss the topological properties of the Hamiltonian \eqref{equ:Hamiltonian}.
Especially the state propagating along the diagonal potential for specific $V^{(n)}_\mathrm{dia, 0}$ (equation \ref{eq:prob_zeros}) seems peculiar as the diagonal is not a boundary between topologically different bulks and is therefore not expected to host any edge states. However, we argue that at these $V^{(n)}_\mathrm{dia, 0}$ the diagonal potential decouples the bulk on either side of the diagonal, and the invariant of our Hamiltonian (\ref{equ:Hamiltonian}) without the diagonal potential can describe the edge state.

One might guess that interpreting the diagonal potential as a defect, the state along the diagonal is, in fact, a defect edge state. We explicitly start from the winding number defined on defect Hamiltonians to show that this cannot be the case.

The Altland-Zirnbauer classification~\cite{altland_nonstandard_1997} has been adopted to Floquet systems with defects~\cite{yao_topological_2017}.
The overall structure of the periodic table of static topological insulators/superconductors remains the same.
However, the Floquet invariants are now winding numbers associated with the band gaps at quasi-energy $\epsilon$ of the time evolution operator $U(\vec{k},\vec{r},t)$, where $\vec{r}$ describes the positional dependence of the time evolution operator.
Notably, because the quasi-energy spectrum is periodic, there is an additional bandgap at $\varepsilon T = \pm \pi$.
More specifically, because the time evolution operator is generally not periodic in time, one can calculate the Floquet invariant through a periodized time evolution operator $U_\varepsilon$~\cite{yao_topological_2017}:
\begin{equation}
U_\epsilon(\vec{k}, \vec{r},t) = U(\vec{k}, \vec{r}, t) \exp\left(it H_{\text{eff}, \varepsilon}(\vec{k}, \vec{r})\right)
\end{equation}
with
\begin{equation}
H_{\text{eff}, \varepsilon}(\vec{k}, \vec{r}) = \frac{i}{T} \ln_{-\varepsilon} (U(\vec{k}, \vec{r},T))
\end{equation}

The subscript epsilon denotes the branch cut of the complex logarithm so that $\ln_{-\varepsilon}(e^{i\phi}) = i\phi$ for $-\varepsilon T - 2\pi < \phi < -\varepsilon T$.
Setting it to a quasi-energy value inside the band gap ensures $U_\varepsilon$ is a continuous function.

The topological classification in each symmetry class is then only dependent on the difference between the spatial dimension $d_s$ and the defect dimension $d_d$.
The dimension of the defect is the dimension of the circle $S^{d_d}$ enclosing the defect.
For example, a point defect in 3D can be enclosed by a 3D sphere $S^3$.
A line defect in 2D (such as the modified diagonal in our system) is enclosed by $S^0$, which are only two points ($\pm 1$) in the plane on either side of the defect.

Our Hamiltonian and its corresponding time evolution operator $U_\varepsilon$ do not possess time reversal, charge conjugation, or chiral symmetry, nor do we enforce any other symmetries (e.g., crystal symmetries).
Therefore, the system belongs to the complex symmetry class A, for which topologically non-trivial phases exist if $d_s - d_d$ is even.
The 2D winding number with our $d_d=0$ defect, given by the diagonal potential, is then the difference of the winding numbers of the bulk on the two sides ($\pm 1$) of the defect:
\begin{equation}
W(U_\varepsilon) = W(U_\epsilon(\vec{k}, +1, t)) - W(U_\epsilon(\vec{k}, -1, t))
\label{equ:App_Winding_Number_Difference}
\end{equation}
where
\begin{equation}
\begin{aligned}
W(U_\epsilon(\vec{k},\pm 1,t))=&\frac{1}{8\pi^2}\int _{\text{BZ}\times S^1}\di k_x \di k_y \di t \times \\
&\Tr((U_\epsilon^{-1} \partial_t U_\epsilon)[U_\epsilon^{-1} \partial_{k_x} U_\epsilon, U_\epsilon^{-1} \partial_{k_y} U_\epsilon])
\end{aligned}
\label{equ:2DClassAWindingNumber}
\end{equation}
Here, the square bracket inside the trace denotes the commutator.
In our system, there is only a band gap at $\varepsilon T = \pi$, as (in contrast to \cite{rudner_anomalous_2013}) we do not have any bipartite on-site potential in the bulk.
We find the winding number
\begin{equation}
W(U_\pi(\vec{k},\pm 1,t)) =
\begin{cases}
1 & \pi < J T < 6.9 \pi \\
0 & \mathrm{else}

\end{cases}
\end{equation}
identical to the one found by Rudner {\em et al.}~\cite{rudner_anomalous_2013} even without the bipartite on-site potential.
Because we chose $J T = 2 \pi$ in Sec.~\ref{sec:system}, $W(U_\pi(\vec{k},\pm 1,t))=1$.

As the bulk Hamiltonian on both sides of the diagonal potential is equal, we see from \eqref{equ:App_Winding_Number_Difference} that the observed edge state cannot be a defect edge state.
Still, each bulk on its own is topologically non-trivial with our driving parameters.
As we have shown in Sec.~\ref{sec:hopping_probabilities}, at $V^{(n)}_{dia,0}$ the probability of the state to hop onto, and hence over the diagonal is zero after a full period $T$.
Therefore, we interpret a quantum state's inaccessibility to the other side as an isolation of each bulk to an effective system with open boundary conditions.
For this system the edge states at the boundary between the bulk to the vacuum is then clearly protected by the winding number $W=1$.

\section{Tracking States}
\label{sec:tracking}
We follow states through changing parameters, e.g., in the closed interval $[V_\mathrm{dia, min}, V_\mathrm{dia, max}]$.
As a starting point, we calculate the Floquet energies $\varepsilon_{0,j}$ and states $\psi_{0,j}$ ($j = 1, 2, \ldots, N_x N_y$) at $V_0 = V_\mathrm{dia, min}$.
Then we loop through the following procedure:
We increase the diagonal potential to $V_i = V_{i-1} + \delta$, with $0 < \delta \le V_\mathrm{dia, max} - V_\mathrm{dia, min}$, and calculate the new Floquet energies $\varepsilon_{i,k}$ and states $\psi_{i,k}$ ($k = 1, 2, \ldots, N_x N_y$).
We compare each of the states $\psi_{i,k}$ to all states $\psi_{i-1,j}$ from the previous step by calculating their overlap (scalar product).
The index $l$ of the most similar previous state $\psi_{i-1,l}$ (with $|\psi_{i,k} \cdot \psi_{i-1,l}| \ge |\psi_{i,k} \cdot \psi_{i-1,j}|\ \forall j$) is stored in a similarity variable $s_{i, k} = l$.

\begin{figure}[hbtp]
\begin{minipage}[t]{0.08\columnwidth}\vspace{0pt}
(a)
\end{minipage}
\begin{minipage}[t]{0.9\columnwidth}\vspace{0pt}
\includegraphics{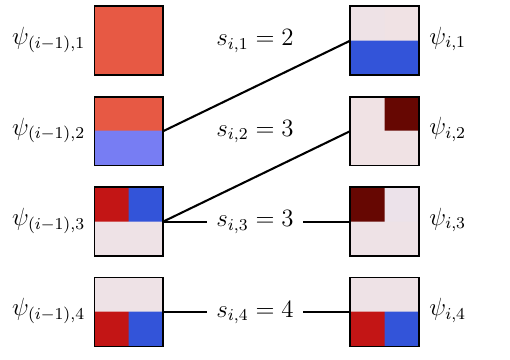}
\end{minipage}
\vspace{2pt}
\hrule
\vspace{6pt}
\begin{minipage}[t]{0.08\columnwidth}\vspace{0pt}
(b)
\end{minipage}
\begin{minipage}[t]{0.9\columnwidth}\vspace{0pt}
\includegraphics{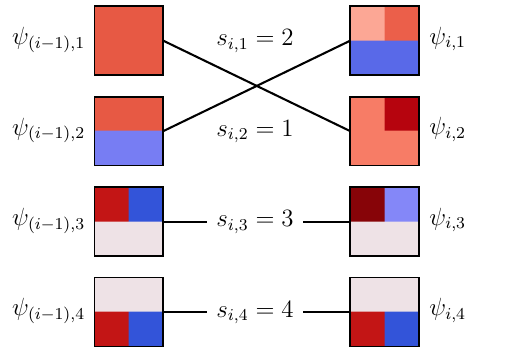}
\end{minipage}
\caption{
a) Example of tracking algorithm failure.
Both $\psi_{i, 2}$ and $\psi_{i, 3}$ are most similar to (i.e. have the largest overlap with) the same state $\psi_{(i-1), 3}$ from the previous step ($s_{i, 2} = s_{i, 3} = 3$).
We need to unambiguously track the evolution of $\psi_{(i-1), 3}$.
Therefore, we must repeat the step from $i-1$ to $i$ using a smaller step size.
b) Example of tracking algorithm success.
The states $\psi_{(i-1), j}$ map biuniquely to the states $\psi_{i, k}$ ($j, k = 1, 2, 3, 4$).
}
\label{fig:tracking}
\end{figure}

If the mapping of states from the previous parameter $V_{i-1}$ to the current parameter $V_i$ is not bijective, i.e., $s_{i, k} = s_{i, k'}$ for $k \ne k'$ (as shown in \autoref{fig:tracking}a), we can not follow the states because they have changed too much.
In this case, $\delta$ must be decreased (in our calculations to $\delta' = \delta / 10$), and the step repeated for $V_i = V_{i-1} + \delta'$.

If the mapping was successful (\autoref{fig:tracking}b), $\delta$ may be kept constant or increased before the next step to $V_{i+1}$.
We repeat these steps until we reach $V_\mathrm{dia, max}$.

This adaptive-stepsize tracking works for non-zero diagonal potentials but breaks down at $V_\mathrm{dia} = 0$, where the doublon states all collapse to localized bulk states at $\varepsilon_\mathrm{F} = 0$ and can not be followed.

\bibliography{references.bib}

\end{document}